\documentclass{IEEEtran}
\usepackage{cite}
\usepackage{amsmath,amssymb,amsfonts}
\usepackage{algorithmic}
\usepackage{textcomp}
\usepackage{graphicx}
\usepackage{xcolor,colortbl}
\usepackage{color,soul}

\usepackage[dvips]{epsfig}

\usepackage[caption=false,font=footnotesize]{subfig}

\usepackage{url,array,multirow,booktabs}
\usepackage{siunitx}
\usepackage{lipsum}
\usepackage[mathscr]{euscript}
\usepackage{comment}
\usepackage{xcolor,colortbl}
\usepackage{color,soul}
\usepackage[utf8]{inputenc}
\usepackage[english]{babel} 
\usepackage{mwe}

\usepackage{diagbox} 
\usepackage{kotex} 
\usepackage[flushleft]{threeparttable} 
\usepackage{lscape} 
\usepackage{tabularx} 
\usepackage{tabu} 

\makeatletter
\newcommand{\thickhline}{%
    \noalign {\ifnum 0=`}\fi \hrule height 1pt
    \futurelet \reserved@a \@xhline
}
\newcolumntype{"}{@{\hskip\tabcolsep\vrule width 1pt\hskip\tabcolsep}}
\makeatother

\def\BibTeX{{\rm B\kern-.05em{\sc i\kern-.025em b}\kern-.08em
    T\kern-.1667em\lower.7ex\hbox{E}\kern-.125emX}}
\begin{document}
\title{Near-Field Challenges in Ultra-Wideband ISAC: Beamforming Strategies and System Insights}
{
\author{Yonghwi Kim, \IEEEmembership{Graduate Student~Member,~IEEE},  Sang-Hyun Park, \IEEEmembership{Member, IEEE}, \\Siyun Yang, \IEEEmembership{Graduate Student Member, IEEE}, Kai-Kit Wong, \IEEEmembership{Fellow, IEEE},\\ Linglong Dai, \IEEEmembership{Fellow, IEEE}, and Chan-Byoung~Chae,~\IEEEmembership{Fellow, IEEE}
\vspace{-10pt}
\thanks{Y. Kim,  S.-H. Park, S. Yang, C.-B. Chae (corresponding author) are with the School of Integrated Technology, Yonsei University, Seoul, 03722 Korea (e-mail: \{eric$\textunderscore$kim, williampark, siyun.yang, cbchae\}@yonsei.ac.kr). K.-K. Wong is with the Dept. of Electronic $\&$ Electrical Engineering, University College London WC1E 7JE, UK. (e-mail: kai-kit.wong@ucl.ac.uk). He is also affiliated with Yonsei Frontier Lab., Yonsei University, Seoul 03722, Korea.  L.~Dai is with the Dept. of Electronic Engineering, Tsinghua University, Beijing 100084, China (e-mail: daill@tsinghua.edu.cn).}
} 
\maketitle

\begin{abstract}
The shift toward sixth-generation (6G) wireless networks places integrated sensing and communications (ISAC) at the core of future applications such as autonomous driving, extended reality, and smart manufacturing. However, the combination of large antenna arrays and ultra-wide bandwidths brings near-field propagation effects and beam squint to the forefront, fundamentally challenging traditional far-field designs. Time-delay units (TTDs) offer a potential solution, but their cost and hardware complexity limit scalability. In this article, we present practical beamforming strategies for near-field ultra-wideband ISAC systems. We explore codebook designs across analog and digital domains that mitigate beam squint, ensure reliable user coverage, and enhance sensing accuracy. We further validate these approaches through large-scale system-level simulations, including 3D map-based evaluations that reflect real-world urban environments. Our results demonstrate how carefully designed beamforming can balance communication throughput with sensing performance, achieving reliable coverage and efficient resource use even under severe near-field conditions. We conclude by highlighting open challenges in hardware, algorithms, and system integration, pointing toward research directions that will shape the deployment of 6G-ready ISAC networks.

\end{abstract}


\begin{IEEEkeywords}
ISAC, near-field, beam squint, mmWave, sub-terahertz, hybrid beamforming, and 3D ray-tracing.
\end{IEEEkeywords}
\vspace{-5pt}
\section{Introduction}
\label{sec:introduction}

\IEEEPARstart{I}{ntegrated} sensing and communication (ISAC) has emerged as a key technology for sixth-generation (6G) wireless networks~\cite{CST_JCAS,JCAS_ISAC}. By enabling communication and sensing to share hardware resources, ISAC reduces signaling overhead, improves spectrum utilization, and enhances performance for both functions~\cite{JSTSP_Optimal_ISAC}. This integration supports a wide range of applications, including real-time detection of vehicles, pedestrians, and obstacles for autonomous systems, sensing-as-a-service in smart environments, and enhanced communication through sensing-aware optimization~\cite{JCAS_ISAC}.

In wireless propagation, especially in the millimeter-wave ($30-300$~GHz) and sub-terahertz ($0.1-1$~THz) bands used to support wider bandwidths~\cite{BM6G,JSAC_BQ_and_TTD}, the near-field (NF) region refers to the space within the arrays Fraunhofer distance, where planar-wave approximations break down~\cite{TCOMM_NF,infocus,NFpolarCB}. Unlike the far-field (FF), where beam direction depends only on angle of arrival or departure, NF propagation couples angle and distance, requiring beamfocusing in both domains~\cite{NFpolarCB}. This enables precise spatial targeting and higher gains but demands more advanced beam training and codebook design than conventional systems~\cite{BM6G}.

For ISAC, NF operation faces additional challenges under ultra-wideband transmission as shown in Fig.~\ref{Fig1}. Wideband signaling exacerbates beam squint, in which the beam direction varies with frequency, degrading performance across the band~\cite{JSAC_BQ_and_TTD}. A common analog domain countermeasure is the use of true-time delay (TTD) circuits; however, in ELAA systems, TTD implementation entails significant hardware cost, power consumption, and complexity~\cite{JSAC_BQ_and_TTD,infocus}. These limitations make TTD impractical for cost-sensitive deployments, highlighting the need for TTD-free NF ISAC solutions that still address both spherical-wavefront effects and beam squint.

In the context of ISAC, the NF regime also offers clear benefits: sensing can directly assist communication by providing accurate user location, thereby reducing beam-training overhead and improving link reliability~\cite{rethinking,curseElbir,unlocking}. In practical multiple-input multiple-output (MIMO) systems, hybrid beamforming with codebook-based beam search is the prevailing operational mode due to its scalability and hardware efficiency~\cite{BM6G,FR22}. However, most existing codebooks are rooted in FF assumptions and do not fully account for NF angleâdistance coupling or the impact of wideband beam squint. Addressing these effects is crucial to realize the full potential of NF ISAC, which calls for codebook designs that are NF-aware, robust to wideband operation, and compatible with TTD-free hybrid beamforming architectures.

\begin{figure*}[t]
  \centerline{\resizebox{1.7\columnwidth}{!}{\includegraphics{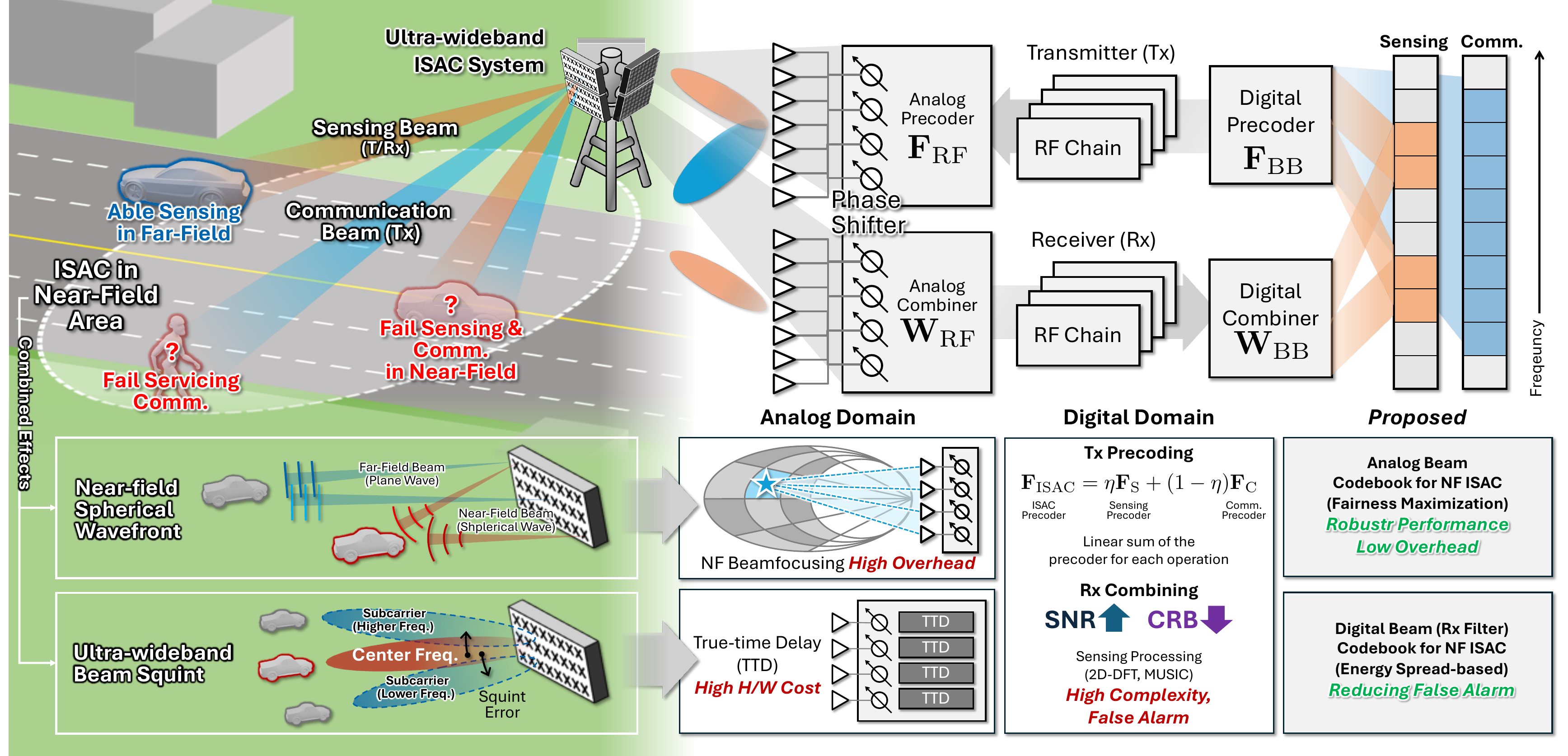}}}
  \vspace{-5pt}
   \caption{An ISAC system in the near-field area based on hybrid beamforming architecture. The conventional methods, such as beamfocusing codebook, and true-time-delay (TTD), induces considerable operation cost. In this article, we introduce some efficient design based on analog domain and digital domain filter codebook.}
   \label{Fig1}
\vspace{-10pt}
\end{figure*}

Recent studies have examined NF propagation and beam squint effects in ISAC, but important gaps remain. In practical ISAC systems with ELAAs, sensing must also operate in a codebook-based manner to align with communication procedures~\cite{BM6G,NFpolarCB}, yet NF ISAC designs built on these paradigms have been insufficiently explored. Some works investigate ISAC beam squint under FF conditions without fully capturing NF spherical wavefront effects~\cite{curseElbir}. Others overlook codebook-based ISAC scenarios~\cite{unlocking} and handle Doppler frequencies without beamforming~\cite{rethinking}. While digital domain methods have also been proposed~\cite{JSTSP_Optimal_ISAC, OptISAC}, they often neglect codebook-based frameworks and assume full channel state information (CSI). As a result, NF ISAC codebook designs that satisfy wideband, hybrid beamforming, and TTD-free constraints remain largely unexplored. 

In this article, we propose a codebook design framework for NF ISAC within hybrid beamforming architectures. Unlike traditional analog approaches, our method eliminates the need for expensive TTD hardware while effectively addressing NF spherical-wavefront and beam squint effects. The key contributions are as follows:
\begin{itemize}
\item We analyze the impact of analog codebooks on NF ISAC and propose a fairness-optimized design that reduces overhead while improving detection probability without complex TTD circuits.
\item We introduce a digital filter-based codebook approach that using analog beamformed information and digitally refines distance accuracy. Using an energy spread (ES) metric, the method reduces false alarms.
\item We validate the proposed framework through 3D ray-tracing simulations, demonstrating substantial improvements in both sensing accuracy and communication performance compared to conventional ultra-wideband analog-centric schemes.
\end{itemize}

The remainder of this article is organized as follows. Section~\ref{Sec.2} describes the combined challenges of NF ISAC such as combined NF region and ultra-wideband. Section~\ref{Sec.3} introduces the analog beam codebook design NF ISAC. Section~\ref{Sec.4} introduces the digital beamforming for control sensing errors. In Section~\ref{Sec.5}, we present the results and insights gathered from the 3D ray-tracing system level simulations. We conclude in Section~\ref{Sec.6} and Section~\ref{Sec.Conclusion} with a summary of our contributions and suggestions for future research.

\begin{figure*}[t]
	\centering
	{{\resizebox{1.5\columnwidth}{!}{\includegraphics{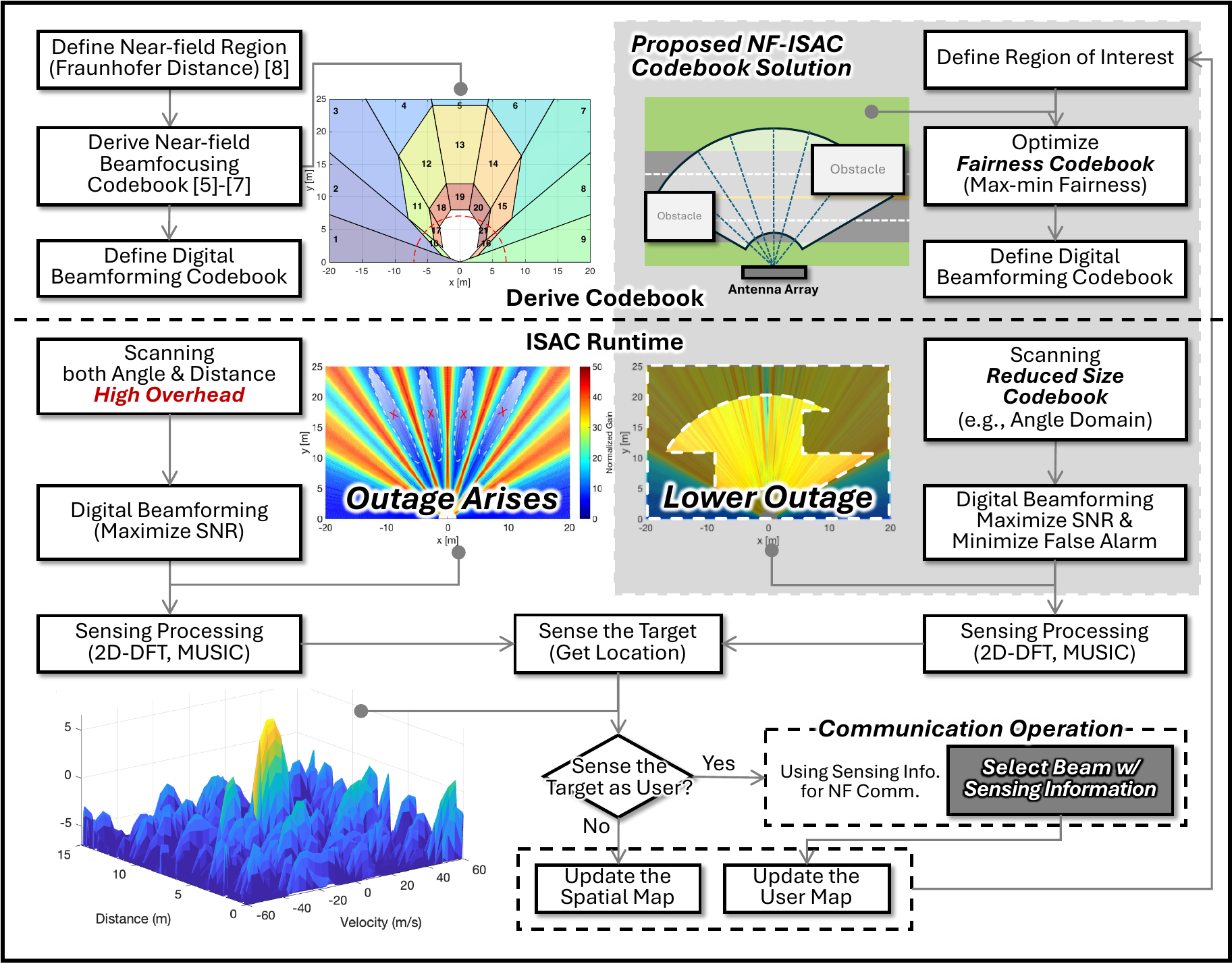}}}}
    \vspace{-5pt}
    	\caption{Illustration of near-field ISAC operation using codebook-based hybrid beamforming. Communication requires beamfocusing using angle and distance, while sensing scans target locations in the near-field. Estimated distance from sensing feedback can improve communication beam alignment, enabling joint optimization with reduced overhead.}
 	\label{Fig2}
	\vspace{-10pt}
\end{figure*}

\section{Challenges in Wideband Near-field ISAC: Hybrid Beamforming-based Approach}
\label{Sec.2}
This section first outlines the NF ISAC system architecture, then describes the propagation characteristics that define its operational challenges, and finally presents two key sensing performance metrics: detection probability and false alarm probability.

\subsection{System Blueprint: How Near-field ISAC Forms and Processes Beams}
As illustrated in Fig.~\ref{Fig1}, NF ISAC employs hybrid beamforming to overcome the severe path loss of high-frequency NF channels. In the analog domain, phase-shifter-based beamforming focuses energy in both communication and sensing directions. The transmitter (Tx) forms multiple beams in parallel~\cite{CST_JCAS, JSTSP_Optimal_ISAC}, while the receiver (Rx) aligns its sensing beam to the corresponding Tx direction. Beam sweeping, implemented via a predefined codebook~\cite{BM6G}, is essential not only for link establishment but also for sensing, where scanning across candidate beams reveals spatial features of the environment~\cite{CST_JCAS}.

Operating with multi-carrier waveforms such as orthogonal frequency division multiplexing (OFDM), the system can dynamically allocate power across beams and subcarriers to adapt between high-throughput communication and high-accuracy sensing~\cite{JSTSP_Optimal_ISAC, OptISAC}. Full-duplex (FD) operation enables simultaneous transmission and reception, capturing reflections from sensing targets or the surrounding environment~\cite{mcISAC}. By correlating transmitted and received signals, the system obtains channel responses containing environmental information~\cite{FDOFDMradar}. Processing these responses with algorithms such as two-dimensional discrete Fourier transform (2D-DFT) or multiple signal classification (MUSIC) yields precise rangeâvelocity profiles, as shown in Fig.~\ref{Fig2}, enabling tight integration of sensing and communication.

\subsection{Nature of Near-field ISAC: Spherical Wavefront and Beam Squint}
The NF region, defined by the Fraunhofer distance~\cite{NFpolarCB}, extends beyond the reactive region for locations close to the array~\cite{NFpolarCB}. In high-frequency systems, an ultra-wideband condition typically refers to a beam with an operating bandwidth exceeding about 10\% of the carrier frequency, which intensifies NF-specific effects such as beam squint~\cite{infocus,curseElbir}.

\subsubsection{Spherical wavefronts}
NF propagation requires beamfocusing in both angle and range, as element-to-user distances vary and must be explicitly modeled~\cite{TCOMM_NF}. This added range dimension increases codebook complexity compared to FF codebook designs, which mainly partition the angular domain.

\subsubsection{Beam squint effect}
Beam squint arises when phase shifters apply frequency-independent phase shifts, ignoring wavelength variation~\cite{curseElbir}. This misaligns beams across subcarriers in wideband systems. TTDs can correct this but are far more complex than phase shifters, with stringent delay resolution and range requirements~\cite{JSAC_BQ_and_TTD}.

\subsubsection{Joint effect of ultra-wideband and NF region}
Combined NF spherical wavefronts and beam squint cause focal points to shift in angle and distance with frequency~\cite{rethinking,unlocking}. This deforms the main lobe, reduces peak gain, and raises sidelobe levels, as further examined in Section~\ref{Sec.4}. NF- and frequency-aware beamforming designs, therefore, are essential.

\subsection{Key Sensing Performance Metrics for Codebook Design: Detection Probability and False Alarm}
In sensing operations, unlike in communications where link endpoints are fixed, performance must be evaluated over arbitrary target locations. NF characteristics necessitate thorough evaluation across the spatial domain. Two essential radar performance metrics are described below.

\subsubsection{Detection Probability}
Detection probability is the likelihood of correctly detecting a target when it is actually present~\cite{mcISAC}. This depends on whether beamforming aligns with the target location and whether the sensing signalâs signal-to-noise ratio (SNR) exceeds a defined threshold~\cite{FDOFDMradar}. In NF operation, if beamforming is not adapted to the targetâs position, detection probability can vary greatly depending on the codebook design, especially in ELAAs where beams are extremely sharp, as illustrated in Fig.~\ref{Fig2}.

\subsubsection{False Alarm Probability}
A false alarm occurs when sensing indicates the presence of a target that is not actually present. The false alarm probability measures the likelihood of such events within a specified region of interest~\cite{mcISAC}. In NF ISAC, the risk of false alarms increases due to beam squint and other NF propagation effects, which can misalign beams and cause communication resources to be directed toward incorrect locations. Effective false alarm management is therefore essential to maintain both sensing reliability and communication efficiency.

\begin{figure*}[t]
\centering
{{\resizebox{1.7\columnwidth}{!}{\includegraphics{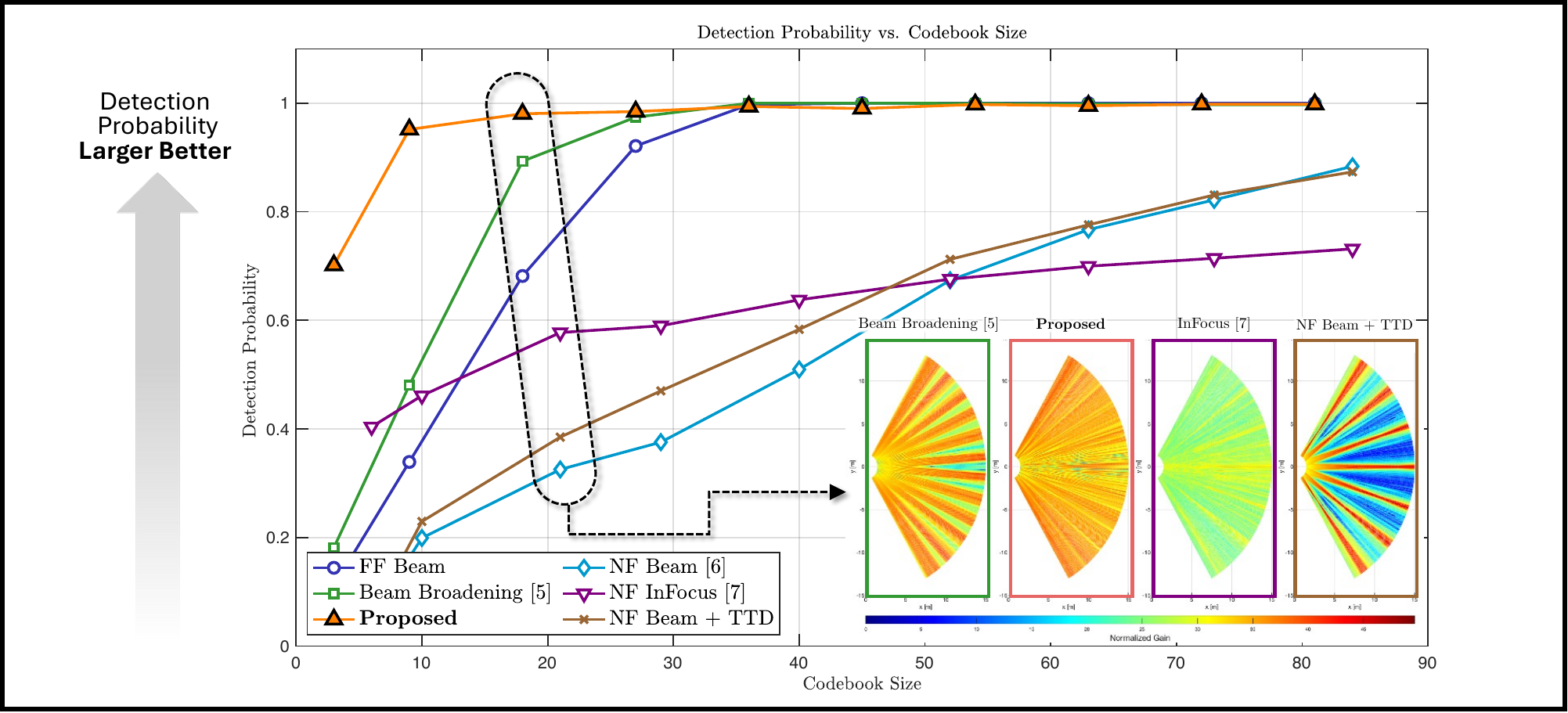}}}}
\vspace{-5pt}
\caption{Detection probability versus codebook size for NF ISAC with $f_c=60$ GHz, $B=5$ GHz, and region of interest (ROI) $\theta\in[-60^\circ,60^\circ]$, $r\in[7,15]$ m. Angle-only codebooks yield higher fairness than angleâdistance designs at comparable sizes, while the proposed method achieves the highest overall detection probability and most robust spatial coverage across the ROI.} 
\vspace{-10pt}
\label{Fig3}
\end{figure*} 

\section{Analog Beam Codebook Design: Pushing Near-field ISAC Detection Performance Without Distance-Domain Overhead}
\label{Sec.3}

\subsection{Problem Statement: Is a Distance-Domain Codebook Really Necessary?}
We investigate how analog beamforming codebook design can enhance ISAC performance in NF region. As shown in Fig.~\ref{Fig2}, we consider an NF beamfocusing codebook and its coverage characteristics. The angular grid spans a $180^\circ$ field-of-view with $20^\circ$ spacing, and the range grid includes $21$ focal distances following~\cite{TCOMM_NF}. The aperture of ELAA is $1$~m with the assumption of uniform linear array (ULA), wwith carrier frequency $f_c = 60$ GHz and bandwidth $B = 5$ GHz, corresponding to the unlicensed band in frequency range 2 (FR2)~\cite{FR22}.

Even with angle- and distance-domain codebooks combined with NF beamfocusing to boost gain, the beams become exceedingly sharp, creating outage pockets (e.g., near boresight) and thus lowering detection probability. These observations suggest two design imperatives for NF ISAC: 1) improve \emph{spatial fairness} over the region of interest (ROI) so that the beam gain remains above a target threshold everywhere, and 2) optimize the codebook primarily in the angular domain, potentially at higher angular density, and leverage frequency diversity as beam squint with digital processing rather than explicit distance-domain beams.

\subsection{State-of-the-Art: TTD-Free Analog Beams for Near-field}
Representative TTD-free analog methods that account for beam squint include \emph{beam broadening}~\cite{JSAC_BQ_and_TTD} and \emph{InFocus}~\cite{infocus}. Beam broadening partitions the array into sub-apertures and points each segment toward the target, effectively reducing the electrical aperture so that the NF region shrinks and squint sensitivity decreases. InFocus equalizes beam gain across subcarriers to improve frequency robustness. Both approaches mitigate squint without TTD hardware, aligning with cost-sensitive ELAA deployments.

\subsection{Our Solution: Fairness-Driven Codebook for Near-field ISAC}
We define an ROI consistent with NF operation and the surrounding environment, partition the region, and optimize \emph{fairness} within each partition to elevate worst-case gain. This sensing-centric view prioritizes spatial partitioning because distance can be estimated via sensing; hence the analog codebook focuses on angle, while distance refinement is handled digitally. Hybrid beamforming with maximum-ratio combining (MRC) across subcarriers maximizes sensing SNR and naturally exploits beam squint diversity.

The optimization enforces unit-modulus constraints on phase shifters via projected gradient descent (PGD). Although maximizing spatial fairness (e.g., maxâ-min gain) is nonconvex, a smooth log-sum approximation enables stable gradient computation and practical codebook learning. This yields consistently higher minimum gain and improved detection probability compared with existing designs.

\begin{figure*}[t]
\centering
{{\resizebox{1.65\columnwidth}{!}{\includegraphics{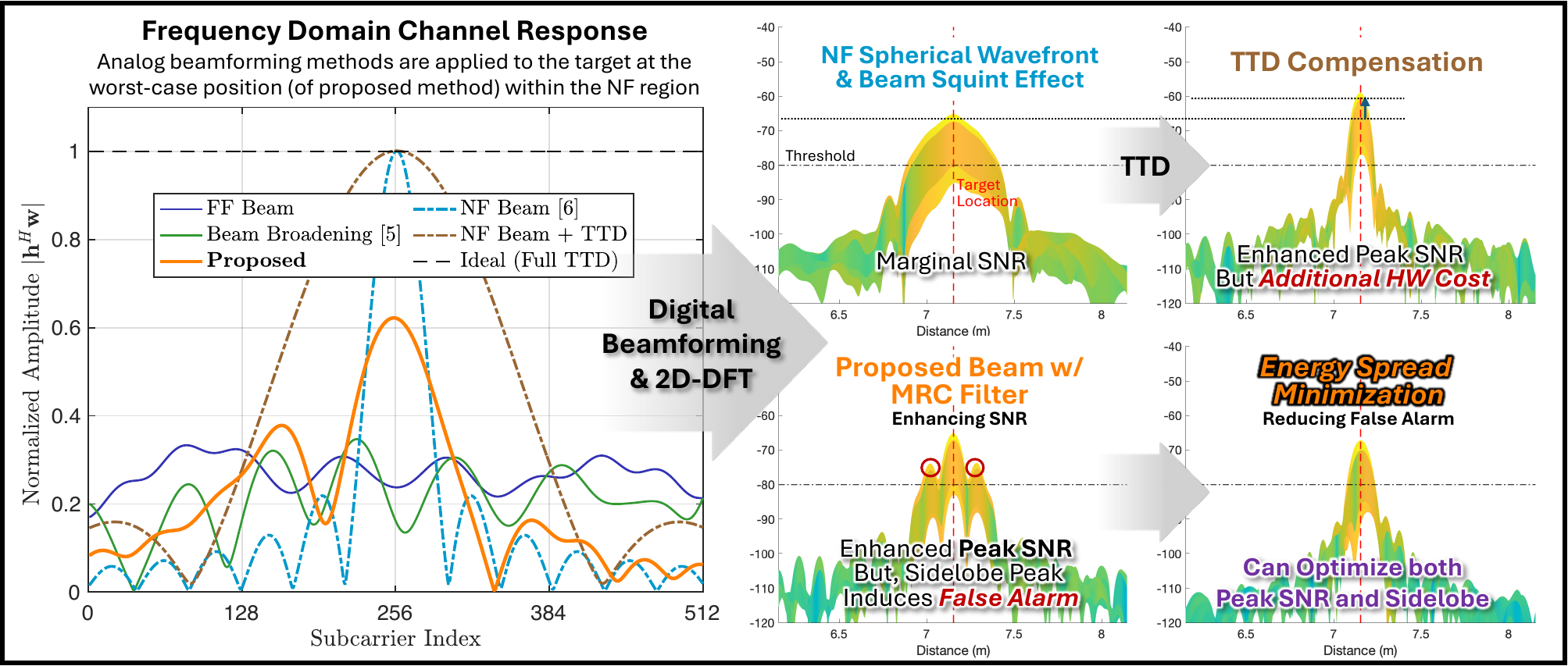}}}}
\vspace{-5pt}
    \caption{(Left) Degradation of the beamformed channel for a near-field (NF) target located at ($7.1$~m, $45^\circ$), evaluated using both NF beamfocusing and proposed codebook under analog beamforming. (Right) Corresponding OFDM radar processing results showing the impact on sensing performance.}
     \label{Fig4}
     \vspace{-10pt}
\end{figure*}	

\subsection{Performance Validation: Beating the Benchmarks in Near-field Conditions}

We simulate an NF channel to evaluate detection probability. The ROI is defined as $\theta \in [-60^\circ, 60^\circ]$ and $r \in [7, 15]$ m, chosen with respect to the reactive NF boundary and OFDM radar constraints, where $\theta$ denotes the angle of arrival and $r$ the radial distance from the array. The ROI is discretized in angle to construct the codebooks, with sizes ranging from $3$ to $90$ beams over $180^\circ$. Detection probability is measured over $10^5$ randomly placed targets as a function of codebook size.

For sensing, we employ $512$ OFDM subcarriers and $256$ OFDM symbols for velocity estimation. To ensure a realistic sensing channel, both line-of-sight (LoS) and non-line-of-sight (NLoS) conditions are considered. Simulations assume a Rician $K$-factor of $30$~dB to model LoS-dominant outdoor links and a phase noise variance of $2^\circ$~\cite{mcISAC}. To simulate the costâperformance tradeoff baseline, the TTD configuration employs a $50$ns resolution with four TTD units\cite{OptISAC}.

Fig.~\ref{Fig3} shows that the proposed method achieves the highest overall detection probability. Angle-only codebooks (e.g., FF beams or beam broadening) outperform angle-distance designs at comparable sizes because they trade peak focusing for superior spatial fairness. Among angle-distance codebooks, InFocus performs well at small sizes; its frequency-domain equalization improves neighboring-user channels under squint. The spatial gain maps in Fig.~\ref{Fig3} (bottom-right) confirm that the proposed codebook delivers the most robust coverage across the ROI.

\section{Digital Beamforming Design: Reducing False Alarms for Hybrid Beamforming}
\label{Sec.4}

\subsection{Combined Channel Degradation from Spherical Wavefront and Beam Squint}
\label{Sec.2.1}

As shown in the leftside of Fig.~\ref{Fig4}, we evaluate the LoS beamformed channel across subcarriers at a worst-performing location for the proposed analog beam, $(7.1~\text{m}, 45^\circ)$. The light blue curve illustrates degradation caused by beam squint, while the brown dotted curve includes partial TTD-based compensation. We adopt a $50$~ns, four-unit TTD setup and an ideal full-TTD baseline across all antennas. The green and blue curves show frequency responses for FF beam, which, despite lower peak gain, remain relatively flat across the band. These results highlight that the combination of NF effects and beam squint produces severe gain loss at frequencies far from the carrier, a key performance limiter for wideband NF ISAC.

\subsection{Conventional Digital Domain Solutions and the False Alarm Problem}

The rightside of Fig.~\ref{Fig4} shows sensing performance of the beamformed channel under combined LoS and NLoS environment. Digital beamforming filters are derived from the simulated LoS channel, and sensing decisions are made by detecting whether 2D-DFT peaks exceed a preset threshold. 

As shown in the bottom-right of Fig.~\ref{Fig4}, MRC achieves peak SNR and beamwidth performance comparable to TTD. MRC maximizes combined SNR across subcarriers, producing a clear delay-domain peak at the target location. At high frequencies, a Cramér–Rao bound (CRB)-minimizing combiner performs similarly to MRC, as channel gain variation across subcarriers dominates over inter-subcarrier delay differences when squint is present. 

However, MRC filter introduces strong delay-domain sidelobes in regions with no targets, substantially increasing false alarm probability. While high peak SNR improves detection probability and reduces estimation error, elevated sidelobes make threshold selection difficult and degrade sensing reliability, particularly in multi-target scenarios~\cite{mcISAC}.



\begin{figure*}[t]
\centering
{{\resizebox{1.65\columnwidth}{!}{\includegraphics{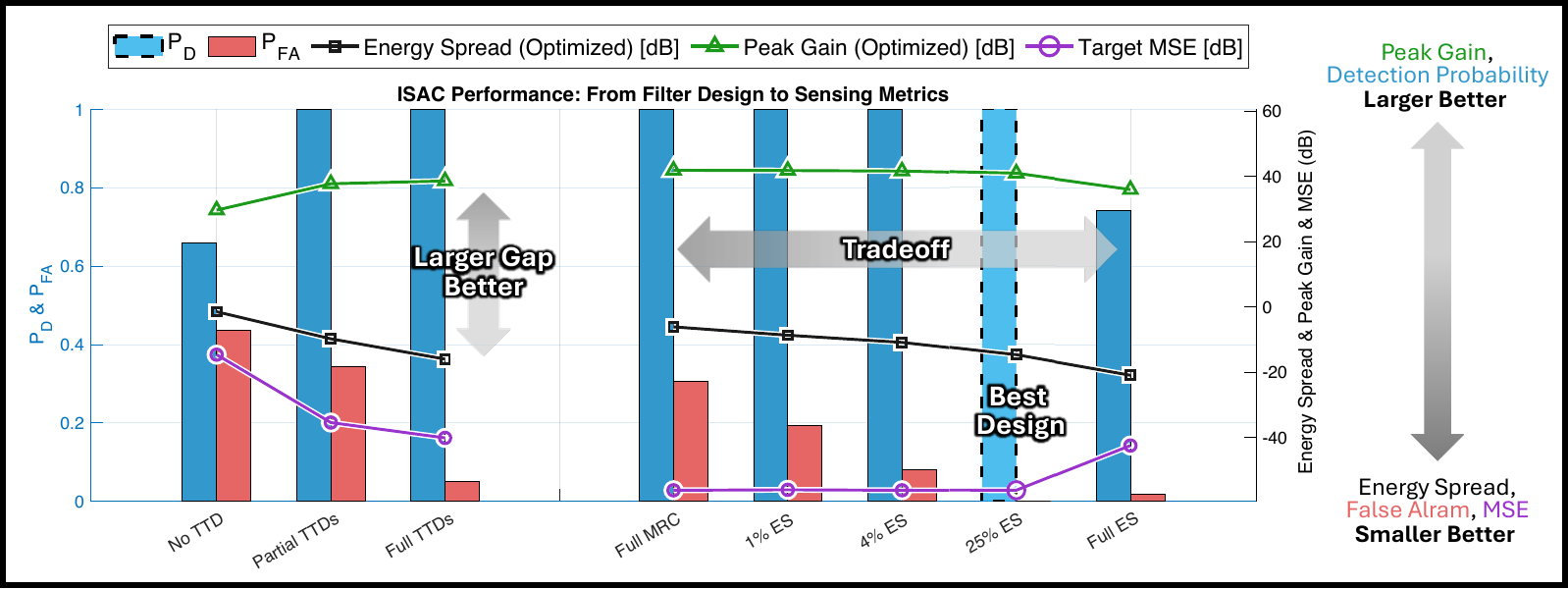}}}}
\vspace{-5pt}
\caption{ ISAC sensing performance under near-field conditions, evaluating detection probability, false alarm rate, and target estimation error. Digital beamforming without TTD improves detection and target MSE, while optimizing the ES metric reduces false alarms to levels comparable with full TTD. The ES and peak SNR metrics offer complementary guidance: SNR improves detection, while ES controls sidelobes and false alarms.} 
\vspace{-10pt}
\label{Fig5}
\end{figure*} 

\subsection{Energy Spread: A Metric for Near-field ISAC Digital Codebook Optimization}
\label{Sec.3.1}

Prior work on sidelobe suppression has largely focused on the angle domain using analog techniques~\cite{CST_JCAS,JCAS_ISAC}. Delay-domain sidelobe suppression under hybrid beamforming and multicarrier conditions remains less explored. Additionally, conventional metrics such as the ambiguity correlation function (ACF) used in OFDM radar design~\cite{mcISAC} are not well-suited to NF ISAC scenarios, where performance is heavily shaped by spatial channel characteristics.

To fill this gap, we introduce the \textit{energy spread (ES)} metric, which jointly captures beamwidth and sidelobe behavior in the delay domain. As shown in Fig.~\ref{Fig4}, a lower ES indicates better focus of signal energy around the true delay, implying improved resolution and fewer sidelobes. This metric is not only a performance indicator but can also be directly minimized through eigenvalue-based optimization. It supports digital codebook design that adapts to NF characteristics, even when analog RF chains only measure angle-domain responses. Peak SNR and ES can be simultaneously optimized through the linear sum of objective functions.

\subsection{Impact of Digital Beamforming Codebook on Sensing Performance}

To demonstrate the benefits of ES-guided design, we evaluate detection probability, false alarm probability, and estimation error, as shown in Fig.~\ref{Fig5}. Simulation results, averaged over $10^3$ trials per location, locations are same as Fig.~\ref{Fig3}, confirm the ES metric's strong correlation with actual sensing performance.

The key findings are as follows. First, applying digital beamforming at the receiver without TTD achieves superior performance in both detection probability and target mean squared error (MSE). Furthermore, when the ES metric is optimized alongside the conventional MRC filter, the system achieves false alarm performance comparable to that of a full TTD setup.
Overall, the ES and peak SNR metrics provide strong theoretical guidance for understanding ISAC sensing performance. Peak SNR primarily governs detection probability and MSE, while ES effectively manages false alarm probability by accounting for sidelobe suppression and delay-domain concentration. Ultimately, ES and MRC represent a tradeoff, and balancing the two allows for enhanced design strategies based on specific NF ISAC requirements.

\section{Conceptual Insights from 3D Ray-tracing-based System-Level Simulations}
\label{Sec.5}

\subsection{End-to-End Near-field ISAC Operation}

The proposed NF ISAC procedure is depicted in Fig.~\ref{Fig2}. Tx initiates sensing by emitting generalized multibeams to simultaneously obtain channel responses while supporting communication. Rx then applies the analog beam from the proposed fairness-optimized codebook for target sensing. In digital domain. the system employs the proposed ES-guided digital beamforming to process sensing in the frequency domain. Once the user or target location is identified through sensing, NF beamfocusing is applied to enhance communication performance by aligning both angle and distance. If a false alarm occurs, communication beamfocusing may be directed toward an incorrect location, which can later be corrected through focused sensing. If the user is not detected, the system falls back to FF beamforming for communication.

\subsection{3D Ray-tracing-based System-Level Simulation Settings}
\label{Sec.4.2}

The system is configured with a $f_c =60$~GHz, $B=5$~GHz, $512$~subcarriers, and a $400$-element ULA. The channel environment is modeled using 3D ray tracing in Wireless InSite, accurately capturing frequency-dependent beam squint characteristics and NLoS propagation effects. We simulate $289$ user locations distributed over a $32~\text{m} \times 32~\text{m}$ area, with a minimum distance of 2~m from the radar. 

\begin{figure*}[t]
	\centering
	{{\resizebox{1.6\columnwidth}{!}{\includegraphics{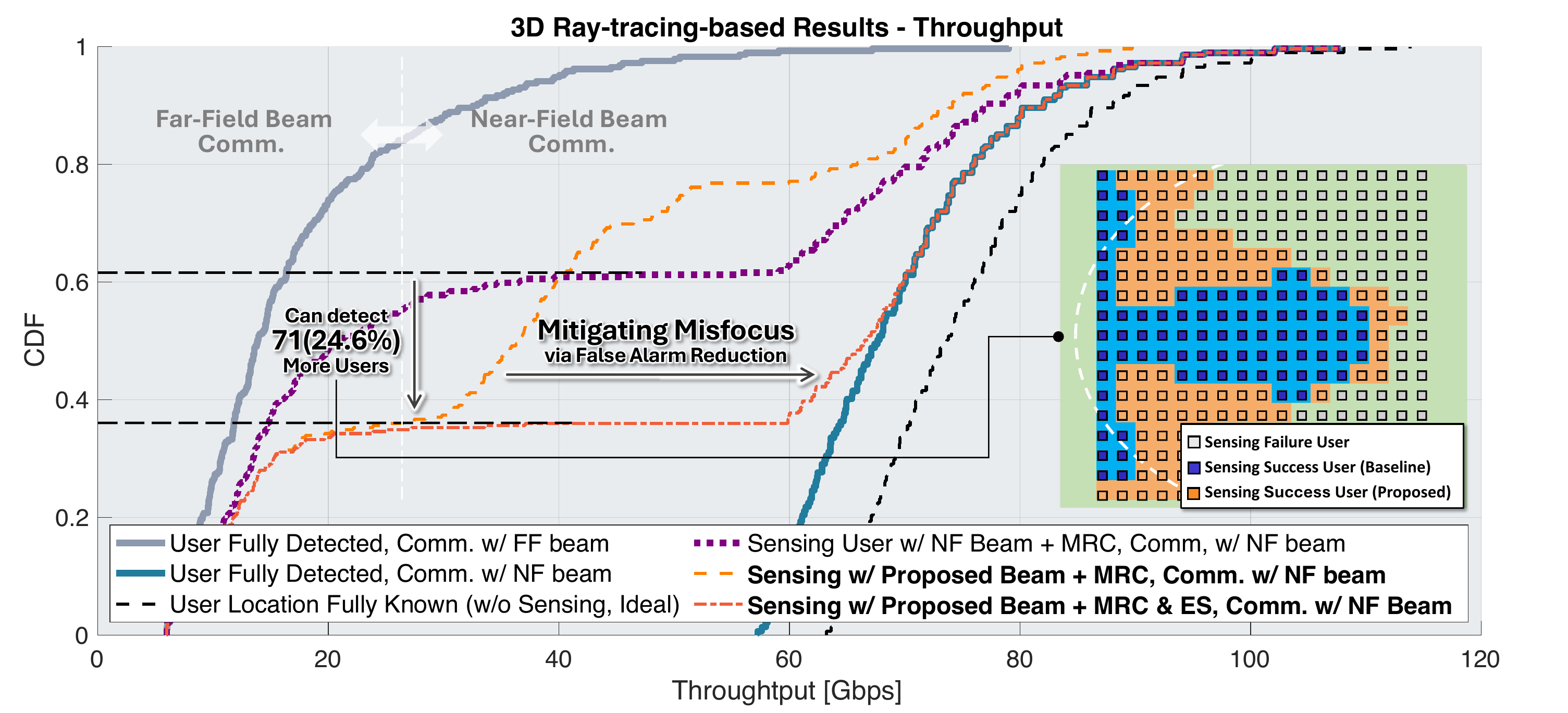}}}}
    	\vspace{-10pt}
 	\caption{User detection and throughput performance in near-field ISAC. NF analog codebook-based sensing (purple) detects fewer users, whereas the proposed codebook (orange) identifies more. Incorporating the proposed digital filter to suppress false alarms further enhances communication throughput by mitigating beam misfocus.}
	\label{Fig6}
	\vspace{-10pt}
\end{figure*}

\subsection{Performance Results: Detection Probability \& Throughput}
\label{Sec.4.3}

Fig.\ref{Fig6} illustrates both user detection coverage and throughput distribution. On the right, blue markers indicate users detected via NF analog codebook-based sensing\cite{TCOMM_NF}, while orange markers show additional users identified using the proposed beam codebook. The white dotted lines mark the NF-FF boundary~\cite{NFpolarCB}.

Using the proposed analog beam codebook (orange line), 163 users are detected compared to 92 with the baseline. Throughput is then evaluated based on these sensing outcomes. The upper bound (green line) represents ideal NF beamfocusing with perfect user detection, while the lower bound (grey line) corresponds to FF beamforming without distance awareness. The proposed digital domain method (red line) narrows this gap by applying a false-alarm reduction filter, mitigating beam misfocus in communication. 

\section{Future Directions and Open Issues for Near-field ISAC}
\label{Sec.6}

\subsection{Site-specific Analog Beam Codebook Design under Spatial Non-Stationarity}
This study employed an analog beam to construct a uniform angle beam codebook. However, this approach may not be optimal across varying user distributions or in environments exhibiting spatial non-stationarity, a condition where the channel characteristics change across the array aperture, especially prevalent in NF regimes. Future directions include adaptive beam codebook designs that leverage optimization or machine learning techniques such as neural networks (NN), reinforcement learning (RL), or model-agnostic meta-learning (MAML) for task adaptation, enabling robust ISAC performance in dynamic or spatially non-uniform conditions.

\subsection{Addressing Hardware Impairments and Nonlinearities}
ISAC performance is highly sensitive to hardware non-idealities~\cite{mcISAC}, particularly in the receiver chain. Phase noise, in-phase and quadrature-phase (IQ) imbalance, and power amplifier (PA) nonlinearities can significantly degrade sensing accuracy. In NF ISAC, the wide bandwidth and high carrier frequency necessitate high-rate sampling and precise hardware calibration. Phase errors, in particular, have a disproportionate impact on delay-based sensing, making them critical to address. Future work must consider accurate impairment modeling and compensation strategies to ensure reliable operation in real-world systems.

\subsection{Evolving Architectures Beyond Hybrid Beamforming}
While hybrid beamforming offers a balance between performance and complexity, emerging hardware paradigms such as dynamic metasurface antennas (DMA), fluid antenna systems (FAS), and movable array architectures present new opportunities. These flexible antenna systems can reduce energy consumption and hardware overhead while enabling spatial reconfigurability. However, they may still be susceptible to frequency-dependent issues like beam squint, particularly in ultra-wideband NF environments. Designing NF ISAC strategies that consider such architectures and developing signal processing techniques that address their unique characteristics will be key enablers for future systems.

\section{{\fontsize{11}{14}\selectfont Conclusion}}
\label{Sec.Conclusion}
In this article, we addressed the critical challenges of near-field propagation and beam squint in ultra-wideband ISAC systems, highlighting why conventional far-field beamforming approaches fall short in 6G scenarios. We introduced practical analog, digital, and hybrid beamforming strategies that not only mitigate these effects but also improve both sensing accuracy and communication reliability. Large-scale simulations, including 3D map-based evaluations, confirmed the effectiveness of the proposed methods under realistic deployment conditions. Looking forward, several open challenges remain. On the hardware side, scalable and cost-effective time-delay elements are essential for practical implementations. On the algorithmic side, new beam management techniques are needed to balance latency, overhead, and performance in highly dynamic environments. From a system perspective, the integration of ISAC into 6G networks requires cross-layer design that jointly optimizes spectrum, computation, and energy efficiency. Ultimately, the near-field nature of 6G ISAC is not just a limitation but also an opportunity: it enables finer spatial resolution for sensing and unprecedented levels of communication reliability. We believe that tackling these challenges will pave the way toward robust, AI-enabled 6G ISAC systems that seamlessly unify connectivity and perception.

\vspace{-10pt}
\bibliographystyle{IEEEtran}
\bibliography{NF_ISAC_hwi_etal}

\end{document}